# Fine-Grained, Language-Based Access Control for Database-Backed Applications

Ezra Zigmond[a], Stephen Chong[a], Christos Dimoulas[b], and Scott Moore[c]

a   Harvard University, USA
b   Northwestern University, USA
c   Galois, Inc., USA

**Abstract   Context**: Database-backed applications often run queries with more authority than necessary. Since programs can access more data than they legitimately need, flaws in security checks at the application level can enable malicious or buggy code to view or modify data in violation of intended access control policies.

**Inquiry**: Although database management systems provide tools to control access to data, these tools are not well-suited for modern applications which often have many users and consist of many different software components. First, databases are unaware of application users, and creating a new database user for each application user is impractical for applications with many users. Second, different components of the same application may require different authority, which would require creating different database users for different software components. Thus, it is difficult to use existing tools to properly limit the authority an application has when executing queries. For this reason, we consider a new, language-based approach to application-specific database security.

**Approach**: Prior work has addressed the difficulty of running applications with least privilege using capability-based security and software contracts, which we adapt to the setting of database-backed applications.

**Knowledge**: This paper's main contribution is the design and implementation of ShillDB, a language for writing secure database-backed applications. ShillDB enables reasoning about database access at the language level through *capabilities*, which limit which database tables a program can access, and *contracts*, which limit what operations a program can perform on those tables. ShillDB contracts are expressed as part of function interfaces, making it easy to specify different access control policies for different components. Contracts act as executable security documentation for ShillDB programs and are enforced by the language runtime. Further, ShillDB provides database access control guarantees independent of (and in addition to) the security mechanisms of the underlying database management system.

**Grounding**: We have implemented a prototype of ShillDB and have used it to implement the backend for a lending library reservation system with contracts for each endpoint to evaluate the performance and usability of ShillDB. Further, we benchmark individual database operations in ShillDB to better understand the language's performance overhead.

**Importance**: Our experience indicates that ShillDB is a practical language for enforcing database access control policies in realistic, multi-user applications and has reasonable performance overhead. ShillDB allows developers to reason about security at the component level, safely compose components, and reuse third-party components with their own application-specific database security policies.

# The Art, Science, and Engineering of Programming

**Fine-Grained, Language-Based Access Control for Database-Backed Applications**

# 1 Introduction

Database-backed applications often require dynamic, fine-grained access control to secure sensitive information. Further, applications may need to control access differently depending on what part of an application is querying the database or on which user's behalf the query is made. Existing techniques make it difficult to meet these security requirements for database-backed applications. This paper introduces SHILLDB, a language that makes it easier to write database-backed applications while enforcing security requirements.

To examine the difficulty of securing database-backed application, consider a student directory application for use by professors at a university. The application is backed by data from two database tables: students, which stores student records, and advising, which maps student IDs to advisors (figure 1).

Suppose the university's policy is that professors can view non-sensitive information about any student (for example, their name and email), but a professor should only be able to view grade-point averages (GPAs) for her own advisees. Even though professors can view some students' GPAs, there are likely parts of the student directory application that should not be able to access any student grades. For example, a component that lets a professor send an email to all her advisees has no reason to access GPAs. If this component cannot access GPAs, then even if a malicious user exploits a bug in this part of the application, she cannot access student grades.

Even this simple web application necessitates controlling database access based on who the logged-in user is and what part of the application is accessing data. Further, the access control policies are specific to the application, as other applications with different policies may use the same database tables.

The following Python snippet shows how a program would typically connect to the database and issue a query. Here, the code uses a popular PostgreSQL adaptor for Python [22] to connect to the database and issues a query to get information for all of the students advised by the logged-in user:

```
1  conn = psycopg2.connect(user="admin", password="12345", host="localhost",
2    port="5432", database="student_records")
3
4  cursor = conn.cursor()
5  cursor.execute(
6    "SELECT * FROM students JOIN advising ON id = student WHERE advisor = %(user)s",
7    {"user": currentUser.getName()})
```

| | students | | | | advising | |
|---|---|---|---|---|---|---|
| *id* | *name* | *email* | *gpa* | | *student* | *advisor* |
| 1 | Mike Birbiglia | birbigs@college.edu | 2.5 | | 1 | Jerome Seinfeld |
| 2 | Tig Notaro | tnotaro@college.edu | 3.9 | | 2 | Jerome Seinfeld |
| 3 | Patton Oswalt | poswalt@college.edu | 3.4 | | 3 | Joan Rivers |

**Figure 1** Schema and example data used by a student directory application.

Ezra Zigmond, Stephen Chong, Christos Dimoulas, and Scott Moore

While this snippet happens to obey the described policy (assuming that the currentUser value is actually the currently logged-in user), a bug in the code (such as forgetting the WHERE clause) could easily violate the policy. How could we enforce that the code obeys the policy? Securing the directory application using access controls at the database level would be difficult or impossible. While commodity database management systems (DBMSs) provide access-control mechanisms, some DBMSs (e. g., MySQL) do not provide the row-level access controls necessary to enforce the access control policies for the application. Further, most DBMS access control mechanisms are based on database users which are cumbersome for large applications because each application-level user must map to a distinct database user. Giving different application components different privilege levels using DBMS-level access controls would require creating even more database users.

Due to the limitations of using DBMS-level access controls for multi-user, multi-component applications, in practice, developers must write security checks as part of application code. However, because most existing database interfaces rely on queries expressed as strings that reference specific table and column names (as seen in the above Python code), enforcing security means inserting checks wherever strings are used in queries. Further, for any third-party code that may access a database, one must read through the code and examine how it constructs query strings to understand what database queries it may run. Because of these difficulties, developers often do not enforce the fine-grained policies they intend and instead settle for more coarse-grained policies, violating the Principle of Least Privilege (POLP). The POLP is the idea that a program should execute with just the authority needed to perform its functionality. When programs violate the POLP and run with more privilege than needed, if malicious users are able to find vulnerabilities in the application's input validation or access control, they may be able to take advantage of this excess privilege.

To address the difficulty of running database-backed applications with least privilege, this paper presents the design and implementation of SHILLDB, a language with a security-focused design that helps application developers control which parts of a database different application components can access (for example, only certain tables) and write fine-grained restrictions on how components can access these parts of the database (for example, restricting that a function can read a table but not update it). Two key pieces of SHILLDB support these goals:

- **CAPQL, the *capability-safe* database runtime of SHILLDB**. Interactions between an application and databases are not directly via query strings but via operations that consume *view capabilities*, an abstraction representing access to a table or a view of a database. View capabilities are unforgeable: a component cannot create them but has to receive them from its calling environment or derive them from other capabilities. CAPQL is a new, capability-based database interface we develop for SHILLDB.
- **The SHILLDB contract language, a language for controlling the database access of SHILLDB applications**. Contracts state and restrict how an application component can use the capabilities it receives. Using contracts, developers can express and enforce application-specific, capability-based security policies. Con-tracts can be added in and refined gradually so programmers can take advantage of the capability safety of SHILLDB even before they write contracts to express finer-grained policies.

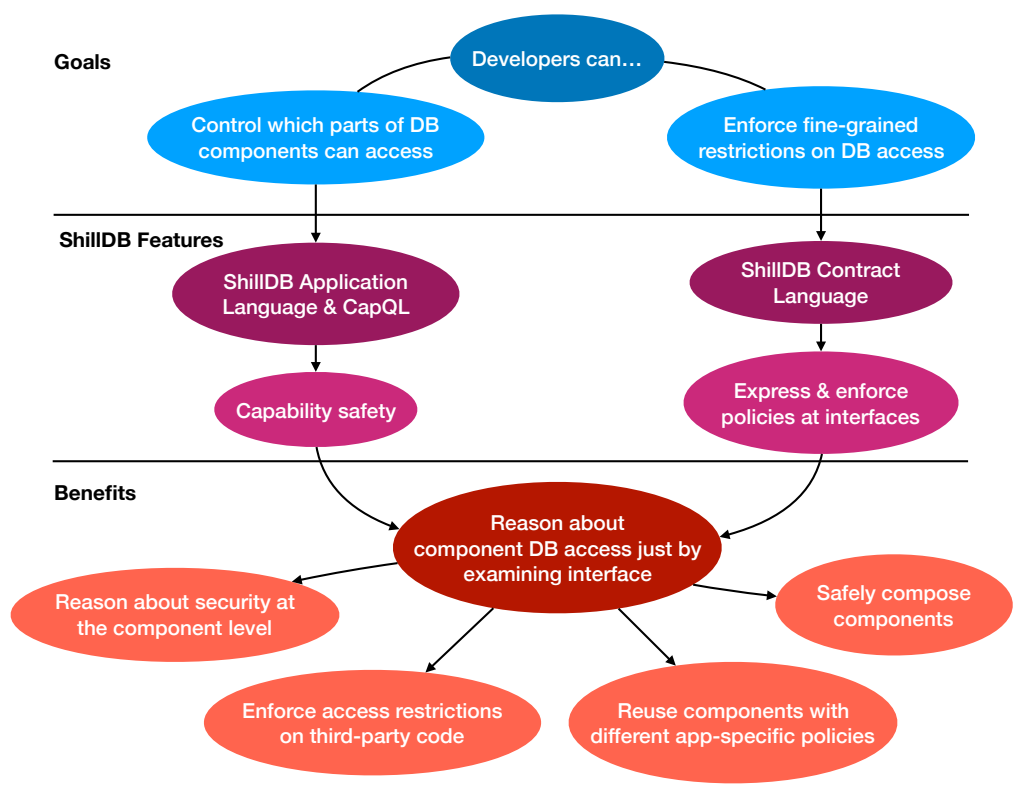

**Figure 2** Overview of how the design goals of SHILLDB map to language features and the benefits of those language features.

Together, these two design points ensure that it is possible to understand the authority a SHILLDB component will have at runtime just by examining what capabilities are passed in and what contracts are part of the component's interface. Since it is possible to reason about and enforce security at the component level, developers using SHILLDB can safely compose components and reuse components with different application-specific policies. SHILLDB makes it easier to reason about the security of third-party components since component consumers control what capabilities are passed to a component. Further, developers can apply their own fine-grained contracts to imported functions. Figure 2 summarizes how the features of SHILLDB relate to its high-level goals and the benefits of the language. These features make it possible to run database-backed applications while following the POLP without needing to rely solely on DBMS-level security tools or brittle application-level security code.

SHILLDB builds on Shill [19], a secure shell scripting language that uses capabilities and contracts to limit access to file system and network resources. Both SHILLDB and Shill are extensions to the Racket programming language [9], and we follow a similar design and implementation to Shill, adding a contract language that is specific for writing database access control policies. Further, while Shill builds on the capability-like standard UNIX file descriptor interface, for SHILLDB, we develop and use CAPQL, a new, capability-based database interface.

Returning to the example of the student directory application, we can consider how the design of SHILLDB helps a developer enforce the application-specific database access control policies. By passing each component only the view capabilities it needs to perform its function, we can be confident that components cannot access arbitrary parts of the database (compare this to the example Python code which can open database connections in any part of the code and access any data the database user can access). For example, while the component that lets professors view student grades needs to access potentially all of the columns in the students table, the component that lets professors email their advisees should not have access to the gpa column. Thus, a developer could pass the email component a capability representing a view of the students table with the gpa column projected away. In section 2, we demonstrate how to create and manipulate view capabilities in SHILLDB.

Further, although the application component for viewing student grades needs access to all of the columns in the students and advising tables, professors should only be able to view their own advisees' grades and should not be able to modify any data. In section 3, we demonstrate how the SHILLDB contract language makes it possible to write these fine-grained access control policies as contracts on view capabilities.

Section 4 describes the implementation of SHILLDB. Section 5 reports on the usability and performance of SHILLDB. Section 6 examines related work. Section 7 concludes. In sum, the contributions of this paper are threefold:

- We introduce CapQL, a new capability-based database interface.
- We introduce SHILLDB which extends the approach introduced by Shill to the setting of database-backed applications and adds contract features for writing fine-grained access control policies on view capabilities.
- We evaluate the usability and performance of the SHILLDB.

**Threat Model**   In SHILLDB, programs written in the capability-safe language are untrusted and are treated as though they may be malicious or contain bugs. A capability-safe SHILLDB program has no access to database resources for which it does not possess a capability and cannot use capabilities in ways that are disallowed by the capabilities' contracts. SHILLDB assumes that contracts on functions are correct insofar as the user executing a SHILLDB program wishes to give the program the authority that the contracts specify. SHILLDB's trusted computing base includes the SHILLDB runtime (and therefore the implementation of Racket), and the implementation of any DBMS that a SHILLDB program accesses. SHILLDB therefore does not defend against programs that exploit flaws in SHILLDB, Racket, or the underlying DBMS. SHILLDB does, however, sanitize untrusted SQL expressions to explicitly defend against programs that seek to exceed their authority through SQL injection. By design, SHILLDB prevents programs from performing any database operation for which they do not have an appropriate capability.

## 2 Design of CapQL

Running database-backed applications with least privilege is difficult when it is not easy to reason about what parts of a database a program can access. In typical database interfaces, queries are written as SQL strings that reference table and column names directly as strings. Typical SQL-based database interfaces make this reasoning hard for two primary reasons. First, queries can refer directly to any tables that the user executing the query has access to, making it challenging to know what tables and rows a program could access without reading through every query the program might issue. Second, queries often conflate different operations into one SELECT statement: a single statement might join tables, select rows, and aggregate data. Because different operations in SQL queries are not always clearly distinct, it is difficult to write fine-grained security policies using the same vocabulary as SQL queries.

To address these problems with SQL, database queries in ShillDB programs use CapQL, a new, capability-based database interface where all interactions with the database are mediated by *view capabilities*. These capabilities are language-level abstractions representing access to a *view* of a database. A view can be thought of as a window into a database that can be queried like a regular table. CapQL provides operations for deriving new view capabilities from existing capabilities (e. g., by projecting away a column) and for fetching/manipulating data in the underlying view. Since all database operations accept only view capabilities (and not string references to table names), it is possible to reason about database authority by restricting the creation and propagation of capabilities. Further, as we will demonstrate in section 3, these operations provide a vocabulary for writing fine-grained security policies

This section presents CapQL and its use in ShillDB as well as an introduction to the ShillDB application language.

### 2.1 CapQL by Example

To introduce CapQL, we demonstrate how we can implement a function in the student directory application to query the database for the GPAs of a professor's advisees. Since the query needs to access the students and advising table, the function must take view capabilities for these tables as arguments (section 2.3 addresses how to create view capabilities in ShillDB). Thus the function definition could be:

```
1 (define (grades-for-advisees v-students v-advising)
2   #| Implementation will go here... |#)
```

The grades-for-advisees function takes two arguments: v-students, and v-advising (view capabilities for the students and advising tables, respectively). Suppose the function should return the name, email, and GPA for all of the logged-in user's advisees. In SQL, the query would look like:

```
1 SELECT name, email, gpa
2 FROM students
```

```
3   JOIN advising ON id = student
4   WHERE advisor = _user -- would be filled in with logged-in user's name
```

To express the same query in CapQL, one can use CapQL *operations* to derive new view capabilities from the given capabilities and ultimately to fetch the data. CapQL provides four primitive operations for deriving new views from existing views: where (which corresponds to selection), select (which corresponds to projection), join, and aggregate (which is like select but allows for aggregations).

The behaviors of where, select, and join operations map closely to that of the corresponding SQL keywords: where takes a view and a WHERE clause and returns a view capability with just the selected rows, select takes a view and a list of column names and returns a view capability that has been projected to contain just those columns, and join takes two views (and, optionally, a WHERE clause) and returns the joined view. Using these operations, we can express the same query in CapQL:

```
1  (select
2    (where
3      (join v-students v-advising "id = student")
4      (sqlformat "advisor = $1" (current-user)))
5    "name, email, gpa")
```

Here, the query joins together the v-students and v-advising capabilities on the condition id = student. From the resulting view, a new view is derived with only the logged-in users' advisees in it (assume that current-user is a function that returns the name of the user for the current login session[1]). sqlformat is a standard library function that safely inserts arguments into a SQL query while avoiding injection attacks. Finally, the view is projected such that the final view has just the name, email, and gpa columns. Note that the select operation supports most of the familiar features of SQL SELECT such as expressions over columns but does not support renaming columns since this could make it difficult to reason about access control policies.

At this point, no queries have actually been executed in the DBMS. In CapQL, creating or deriving a view capability is distinct from fetching the data in that view. Given a view capability, the fetch operation executes the query and fetches results from the DBMS. Adding a fetch call, we can complete the grades-for-advisees function:

```
1  (define (grades-for-advisees v-students v-advising)
2    (fetch
3      (select
4        (where
5          (join v-students v-advising "id = student")
```

---

[1] We imagine current-user as a primitive of the capability-safe and ambient languages of ShillDB. As such, it is part of ShillDB's trusted codebase. The primitive can be configured when launching ShillDB to find the information it needs from a database, a config file or some other source. Thus, the power of the primitive is determined by the security profile of the user that launches ShillDB rather than any untrustworthy capability-safe code.

```
6      (sqlformat "advisor = $1" (current-user)))
7    "name, email, gpa")))
```

### 2.2 Modifying View Data

The grades-for-advisees function involves only fetching data, but CAPQL also provides delete, insert, and update operations that correspond to their SQL counterparts. Like fetch, these three view modification operations consume a view and issue a query to the underlying DBMS.

When inserting into or updating a CAPQL view, we require that the new data must satisfy the WHERE clause in the view definition (corresponding to the WITH CHECK OPTION functionality that many DBMSs provide when creating views). As an example, consider the following update which attempts to modify all student entries with a GPA less or equal to 2.5 to have a GPA of 3.7. This query will fail because it would cause the Mike Birbiglia entry in the table to no longer satisfy the gpa <= 2.5 clause:

```
1  > (define students-with-low-grades (where v-students "gpa <= 2.5"))
2  > (update students-with-low-grades #:set "gpa = 3.7")
```

update: violated view constraint: gpa <= 2.5

To execute the desired update, one can instead provide a WHERE clause argument to the update operation to specify the subset of rows to update. This update will succeed since CAPQL will not treat the WHERE clause as part of the view definition:

```
1  > (update v-students #:set "gpa = 3.7" #:where "gpa <= 2.5")
```

Operations that modify data are not well-defined for all views. For example, if a view contains a non-simple column (e. g., gpa + id), it is not clear how an update to this column should modify the underlying table. CAPQL does not provide a solution to this *view update problem*. Instead, we use conservative restrictions similar to those found in commercial DBMSs [14, 20] to determine which views are insertable and deleteable and which columns are updatable. For example, any view that is the result of a join or an aggregation is not insertable or deleteable and has no updatable columns.

### 2.3 Where do View Capabilities Come From?

While the SHILLDB application language (the *capability-safe language*) cannot create new capabilities directly, it is necessary to use ambient authority (that is, the authority derived from the program's execution context [16]) to create a starting set of capabilities. For this reason, SHILLDB also has an *ambient language* which can use ambient authority but is otherwise highly restricted: ambient programs can only create capabilities for database tables, perform a limited set of actions on capabilities (such as using a where operation to restrict a view), and apply functions from the capability-safe language. These restrictions are so that ambient programs will be short and simple, making it easy to reason about what ambient authority they use. Since the initial set of capabilities must be created in the ambient language, ambient programs serve as the entry point for running ShillDB programs. Note that the ambient program can only access database resources that the user invoking the program can access using any database credentials that might be supplied to the program. In the case of a multi-user application, the capability-safe program may choose to restrict the starting set of capabilities dynamically based on who is accessing the server, as we illustrate in section 3. The split between an ambient language and a capability-safe language follows the design of Shill [19].

In the ambient language, a new view of a single table can be created by supplying the name of the database and the name of the table to the make-view function. The resulting view contains all of the data in the table, analogous to a **SELECT** * SQL query.

Figure 3 contains a complete example of invoking a capability-safe function from an ambient program. Figure 3a contains the grades-for-advisees function in the context of a full ShillDB program. The #lang shilldb/cap annotation on line 2 indicates that this file contains capability-safe ShillDB code. Line 4 exports the function for use by ambient programs or other capability-safe programs. The rest of the code listing is the same grades-for-advisees shown previously.

Figure 3b shows an ambient program that invokes grades-for-advisees. The #lang shilldb/ambient annotation on line 2 indicates that this file is an ambient ShillDB program. Line 4 imports the definition of grades-for-advisees from the capability-safe program from figure 3a. Lines 6-7 define view capabilities that are passed to the capability-safe function on line 9.

## 3   Design of the ShillDB Contract Language

While we have shown how ShillDB controls database access through capabilities, capability safety on its own does not provide guarantees on how capabilities will be used. For example, although the informal specification of the student directory application states that professors should not be able to view GPAs for students who are not their advisees, there is no guarantee that the implementation of grades-for-advisees will use the given view capabilities only in accordance with this policy. In this small example, it is easy enough to examine the implementation and check that it obeys the intended access policy; however, in a large application, it would be much more difficult to decide if the database operations satisfy the intended policies.

To address this difficulty, every function a ShillDB program exports must be accompanied by a *contract* which places restrictions on what arguments can be passed to the function and how those arguments can be used, serving as *executable documentation* for the function's authority. Contracts on view capabilities can specify what *privileges* are required on a capability, where privileges correspond to the authority to invoke a particular CapQL operation. For example, a contract may specify that a view capability passed as an argument has only the +fetch privilege in which case the view can be read but cannot, for example, be used to update the database or be joined with another view.

Figure 4 shows an overview of how contracts and capabilities together control access to database tables in ShillDB. In the figure, the ambient program creates a

```
1  ; cap.rkt
2  #lang shilldb/cap
3
4  (provide grades-for-advisees)
5
6  (define (grades-for-advisees v-students v-advising)
7    (fetch
8     (select
9      (where
10       (join v-students v-advising "id = student")
11       (sqlformat "advisor = $1" (current-user)))
12      "name, email, gpa")))
```

**(a)** Capability-safe program.

```
1  ; ambient.rkt
2  #lang shilldb/ambient
3
4  (require "cap.rkt")
5
6  (define students (make-view "database.db" "students"))
7  (define advising (make-view "database.db" "advising"))
8
9  (grades-for-advisees students advising)
```

**(b)** Ambient program.

**Figure 3** An ambient program creates capabilities and invokes a capability-safe function.

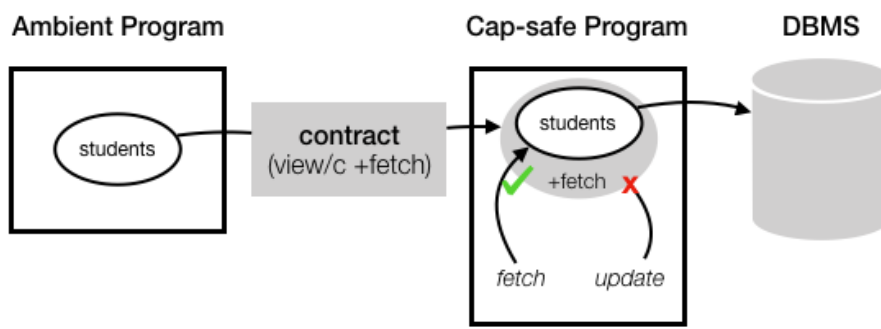

**Figure 4** Contracts and capabilities in SHILLDB.

capability for the students table and passes the capability to a capability-safe program. The interface of the capability-safe program then applies the contract (view/c +fetch) to the given capability which declares that the only allowed operation on the view is fetch. The contract serves as a wrapper or *proxy object* that receives operations that the capability-safe program invokes on the capability. If the program calls fetch on the capability, the contract will forward this operation on to the view capability, which in turn will fetch the contents of the view from the DBMS. If, however, the capability-safe program invokes an operation that is prohibited by the contract (e. g., update), the wrapper will reject the operation and signal a contract failure, blaming (that is, assigning responsibility for the failure to) the capability-safe program. A capability is wrapped in a new proxy object whenever it passes through a contracted function, so a raw view capability can have multiple wrappers enforcing multiple contracts. Contracts are enforced conjunctively since any operation must pass through all of the proxies (while this incurs runtime overhead per wrapper, improving the space and time efficiency of multiple contract wrappers is active area of research [8]).

Since the capability-safe program can access the DBMS only through the given capabilities, contracts mediate all interactions between programs and the DBMS. Capability-safe programs are a collection of functions that all come with contracts and clients of these programs (other capability-safe programs or ambient programs) can only interact with the database by calling these capability-safe functions.

While both contracts and capabilities are used in ShillDB to control access to database tables, they serve distinct purposes. A capability reifies unrestricted access to a database view; contracts wrap around capabilities and restrict how the capability can be used. A contracted capability (i. e., a capability with a contract wrapped around it) can be used wherever a capability can be used. For the remainder of the paper, we use the term "capability" to include raw capabilities (i. e., capabilities without any contracts) as well as contracted capabilities.

The rest of this section introduces the SHILLDB contract language by continuing the student directory application example and demonstrating how to express and enforce the desired database access policy using the contract language. Appendix A contains a reference of SHILLDB contract syntax.

### 3.1 A Simple Contract

Consider a function display-students that displays basic information about all students as well as GPA information for any student who is advised by the logged-in professor (this is similar to the grades-for-advisees function from section 2 except that it will also retrieve some data for non-advisees). This function takes in a view capability for the students table and a view capability for the advising table, so the signature is the same as grades-for-advisees:

```
1 (define (display-students v-students v-advising)
2   #| Implementation goes here |#)
```

In SHILLDB, one can attach a contract to the function to restrict its arguments. A simple contract would prescribe that v-students and v-advising are view capabilities. For simplicity, suppose we are not concerned about the return value of the function.

```
1 (provide
2   [display-students
3    (-> view/c view/c any)])
```

The **provide** form attaches a contract to an exported value (in future snippets, we may omit the **provide** form and just write the contract for brevity). The -> *contract combinator* takes a contract on each of the function arguments (in order) along with a contract on the return value and returns a contract that will check that the argument and result contracts are satisfied. view/c is a built-in contract that checks that a value is a view capability. The **any** contract on the result is a contract that is always satisfied. view/c and **any** are known as *flat contracts*: that is, contracts that can be checked immediately when arguments are passed to the function or when the function returns. This is analogous to standard Racket contracts that check what kind of data functions consume and return such as the integer? and string?.

This contract, like a simple type signature, is sufficient to prevent certain obviously wrong programs (such as a program that invokes display-students with arguments that are not views), but so far does not restrict how the implementation of display-students can use the provided view capabilities.

## 3.2 Privileges on View Capabilities

The implementation of display-students legitimately needs to restrict the provided view capabilities (using select and where), join them together to match students to advisors, and fetch data from the views. However, the function should not modify data in the underlying tables. To express this policy, one can modify the previous contract to specify which privileges are required on the view capabilities (for clarity, the contracts on the view capability arguments are labelled with comments):

```
1 (->
2   (view/c +select +where +join +fetch) ; v-students
3   (view/c +select +where +join +fetch) ; v-advising
4   any)
```

The above SHILLDB contract employs privileges to indicate that the function it protects can only invoke where, join and fetch on its two view capability arguments. The SHILLDB view/c contract combinator can take privileges to specify what operations can be invoked on a view capability. Each privilege (e. g., +select or +join) represents the right to invoke the corresponding CAPQL operation on the capability. Within the function, allowed operations will be proxied to the underlying view capability, while operations for which there is not a corresponding privilege (such as delete in this example) will be rejected by the contract wrapper.

Unlike the earlier simple view/c contract that checks only that the argument is a view, these view/c contracts are *not* flat contracts since we cannot determine at function application time whether the function body will use the view in accordance with the contract. Instead, these contracts wrap the function argument within the function body and intercept operations on the capability to check if they are allowed (as illustrated in figure 4). This is analogous to higher-order function contracts in Racket which wrap function values so that subsequent applications of the function can be monitored for contract violations.

This contract is sufficient to prevent certain buggy or malicious implementations of display-students (such as an implementation that modifies students grades), but it is overly permissive: the contract on v-students includes a +fetch privilege which means that the function can read the GPA of any student, violating the intended policy.

### 3.3 Privilege Modifiers

While the +fetch privilege is too permissive for the desired policy, privileges support *modifiers* which further refine what operations a privilege permits. A complete listing of privilege modifiers is in appendix A.2.

```
(->
  (view/c +join
       [+fetch #:restrict (lambda (v) (select v "name, email"))]
       [+where #:prohibit "gpa"]) ; v-students
  (view/c +select +where +join +fetch) ; v-advising
  any)
```

To enforce that the function body should not be able to freely access the gpa column of the students table, we can use the #:restrict modifier on the +fetch privilege. This modifier takes a function that transforms the view into another view and applies the function before executing the fetch operation. On line 3, the +fetch privilege has been modified so that the gpa column will be projected away before any fetch operations.

To prevent GPA information from being revealed indirectly, two more privilege modifications are needed. On line 4, the #:prohibit modifier on +where prohibits WHERE clauses that mention gpa such as gpa < 3.0 since these queries could be used to guess GPAs by examining which students are in the result set. Note the result of a join inherits the contracts of both joined views, so joining v-students with v-advising could not be used to derive an unmodified +fetch privilege on the gpa column.

Note, however, that this contract is now too restrictive: the function body can not fetch the gpa column for any students, but the intended policy is that professors can view their own advisees GPAs.

### 3.4 Join Groups and Contracts on Join

Intuitively, we wish to express that the function can fetch the gpa column of the students view only after the table has been restricted to just the logged-in user's advisees. This can be thought of as a policy involving joins: if the students and advising tables are equi-joined on student ID (the id and student columns) and the resulting view restricted to the logged-in user's advisees, the resulting view can then have an unrestricted fetch privilege.

While most operations on view capabilities operate only on a single capability, join is inherently *binary*. In many cases, it is natural to think of policies in terms of how a particular set of views can be joined together. To this end, SHILLDB provides the ->/join contract combinator for writing *join group* constraints on function arguments. A join group confers the privilege that any views in the group can be joined together.

Using an ->/join contract, we can use a join group to express the intended policy for the display-students function. An ->/join contract has two parts: the definition of the join group and the contracts on the function arguments and return value.

```
1  (->/join
2    ([X #:post (lambda (v) (where v (sqlformat "student = id AND advisor = $1" (current-user))))
3        #:with (view/c +select +where +fetch)])
4    #| Contracts on the arguments and the result will go here... |#)
```

Lines 2-3 define a join group called X. Line 2 uses the #:post modifier to express that any join on views in the group must join on the two student ID columns and restrict the resulting view to just those rows where the advisor column matches the output of the current-user function (again assume this function returns the name of the user for the current login session).[2] Line 3 uses the #:with modifier to express that the view resulting from joining views in the join group should derive unrestricted +select, +where, and +fetch privileges.

Adding in the contracts on the arguments and the return value from the previous example gives the final contract:

```
1  (->/join
2    ([X #:post (lambda (v) (where v (sqlformat "student = id AND advisor = $1" (current-user))))
3        #:with (view/c +select +where +fetch)])
4    [(view/c +join
5         [+fetch #:restrict (lambda (v) (select v "name, email"))]
6         [+where #:prohibit "gpa"])
7     #:groups X] ; v-students
8    [(view/c +select +where +join +fetch) #:groups X] ; v-advising
9    any)
```

Lines 1-3 are the same as above. Lines 4-9 are the same as the contracts on the view capabilities and the return value from the contract in section 3.3 except that the two view/c contracts have been annotated with #:groups X (lines 7 and 8) to indicate that the views are in the join group defined above. Adding the views to the group modifies the +join privileges such that the #:post and #:with modifiers in the group definition will apply when the two views are joined.

While this contract is sufficient to enforce the desired access control policy, it cannot enforce stronger information flow policies such as "no professor can ever learn the GPA of a student who is not her advisee." For example, if the GPA values returned from this function were passed to another function that had a capability that allowed it to write to the database, the program could write those values somewhere else in the database where other users could access them. For a brief discussion of alternative approaches to database security that deal with information flow, see section 6.

---

[2] Note that the #:post condition in the contract will construct a query dynamically when a join is applied to the contracted capability. If the WHERE clause defined in the contract were invalid, this would go undetected until a join is applied to the capability at which point CapQL will check that the provided SQL clause is valid (see section 4.1).

### 3.5 How are Contracts Enforced?

View contracts work in two different ways. First, they might prohibit an operation outright. Second, they may allow an operation but modify its effect. Simple privileges are an example of the first sort of contract: calling update on a capability without the update privilege will result in a contract failure (as shown in figure 4), and program execution will stop. Similarly, attempting to join a view with another view that is not in the same join group will cause a contract failure.

The second type of contract enforcement is used for contracts with #:restrict modifiers. The modifiers are enforced through *dynamic query rewriting*: any operations invoked on the view are rewritten to enforce the security policy. For example, if a view had a delete privilege with the modifier #:restrict (**lambda** (v) (where v "id = 3")), any delete queries on the view will be rewritten by adding id = 3 to the WHERE clause of the query to enforce that only rows with the ID 3 can be deleted.

## 4 Implementation of SHILLDB

We have implemented a prototype of SHILLDB in Racket [9] using Racket's macro system and tools for creating languages [28]. Racket's macro system allows writing functions from syntax objects to other syntax objects (and is thus a form of source-to-source compilation). SHILLDB syntax is defined using macros that expand into Racket code which the standard Racket runtime can execute. SHILLDB contract combinators like ->/join are compiled into contract combinators from Racket's contract system. Implementing SHILLDB in this way allows reuse of Racket's contract implementation and makes it easy to reuse other Racket features that do not compromise security.

### 4.1 CAPQL

We implemented a prototype of CAPQL in Racket on top of Racket's standard database library [4]. View capabilities are structs that store metadata (such as information about the underlying table's schema) and an abstract syntax tree for the query that the view represents. Operations produce new views from existing views by manipulating the syntax tree. The query is concretized into database-specific SQL syntax only when an operation is invoked that requires database interaction (i. e., fetch, update, delete, or insert). CAPQL views are language-level abstractions: no operations create DBMS-level views. We currently support SQLite3 [29]; other DBMSs can be added modularly.

The update and insert operations use database *triggers* to enforce view constraints. For these operations, we install a trigger on the underlying DBMS table, execute the query, and then remove the trigger. The trigger installed for insert operations checks that each inserted row satisfies the WHERE clause of the view, and aborts the insert if any row does not satisfy it. Similarly, for an update, the installed trigger checks that each updated row still satisfies the WHERE clause of the view, and aborts the entire update if not. Using SQLite's temp trigger functionality [24] (which installs a trigger just for the current database connection), the trigger-based approach works even if multiple applications with different access control policies access a table concurrently. Other DBMSs have similar mechanisms, so this approach is portable.

To provide early detection for invalid operations on views, CapQL parses and validates any user-provided SQL expressions in WHERE clauses, update expressions, and select statements. The SQL parser supports a limited but representative set of SQL expressions (boolean and arithmetic expressions over columns, strings, and number literals). This validation ensures that when an operation is eventually executed, it will produce valid SQL (e. g., all column names are unambiguous and refer to columns in the view). This validation is for usability rather than security: Racket's standard database library provides robust checks against SQL injection, and the underlying DBMS will reject queries that correspond to invalid view operations (e. g., referencing a column that has been projected away). Early detection allows CapQL to raise an error as soon as an invalid view is created, which facilitates debugging.

**Limitations** CapQL does not currently support all features of SQL, such as different kinds of joins, union and intersection operations, or nested queries in WHERE clauses. It also does not provide functionality for running a sequence of queries in a transaction. These limitations are not fundamental, and the features could be added to CapQL.

**4.2 Contracts**

ShillDB contracts are implemented using Racket's contract facilities. Contracts on view capabilities create proxy objects around capabilities, allowing contracts to interpose on operations and check privileges. Contracts also allow for privilege modifiers that can restrict or modify an operation's arguments. We thus implement proxy objects as struct impersonators [27] which allow redirecting or modifying operations on structs. The use of these proxy objects follows the implementation of contracts in Shill [19], but Shill does not use struct impersonators since Shill contracts do not modify operations.

**Join Contracts and Join Groups** Because joins are binary, the join operation requires special consideration in implementation. Each view can store *join constraints* for that view. A join constraint is a special stateful view contract. Adding a view to a join group wraps the view in a join constraint and updates the state of the contracts on all views in the group to include the new view. Invoking join on the view checks if the other argument of the join is in the contract state. The ShillDB ->/join contract combinator is a macro that hides implementation details of join groups from users by creating the join constraints and wrapping views in the join group with the constraints.

### 4.3 Language Restrictions

To provide capability safety, SHILLDB must restrict available Racket functionality. The restrictions in the ambient language have already been discussed in section 2.3, so here we only consider the capability-safe language.

**Capability-safe Language**   The capability-safe language disallows access to Racket standard libraries that could access databases using ambient authority (e. g., the system library or database library). Mutable global variables are not allowed to prevent functions storing capabilities between calls. Capabilities cannot be serialized or deserialized so that programs cannot store capabilities in database tables. Capability-safe programs can import definitions only from other capability-safe programs (and vetted standard libraries) to prevent access to functions that can use ambient authority.

## 5  SHILLDB in Action

To evaluate the usability and performance of SHILLDB, we used it to implement a library reservation system. We also used benchmarks to better understand the performance characteristics of individual operations.

### 5.1 Case Study

We have used SHILLDB to implement a library reservation backend. The server provides five endpoints with the following functionality and high-level security policies:

- **reserve**: Given the ID of a book and a view of the reservations table, reserve that book for the logged-in user. Should not modify any existing reservations or make a reservation for a different user.
- **my-reservations**: Given views of the table of books, authors, and reservations, return details for all of the logged-in user's reservations. Should not modify any data or reveal information about reservations belonging to other users.
- **remove-reservations**: Given the ID of a reservation and a view of the reservations table, delete the given reservation. Should not delete the reservation if it does not belong to the logged-in user.
- **search-author**: Given the name of an author and views of the tables of books and authors, return all the books by that author. Should not modify any data.
- **num-reservations**: Given the ID of a book and a view of the reservations table, return the number of reservations for the book. Should not modify any reservations and should reveal no details of specific reservations.

Appendix B shows the database schema and the declaration of each endpoint with the contract corresponding to the security policies described above. The server is written in a combination of Racket and SHILLDB code. The main loop of the server is implemented in Racket (to take advantage of Racket libraries for web servers which have not been vetted for use in SHILLDB). The backend functionality that interacts with the database is implemented in capability-safe ShillDB code. A short ambient ShillDB program provides an interface between the Racket and ShillDB code and creates the capabilities needed by the capability-safe implementations of the server endpoints. When the server receives a request, it calls the corresponding ShillDB function to handle it. The server implementation required 35 lines of Racket code and 67 lines of capability-safe ShillDB code (of which 21 lines specify contracts).

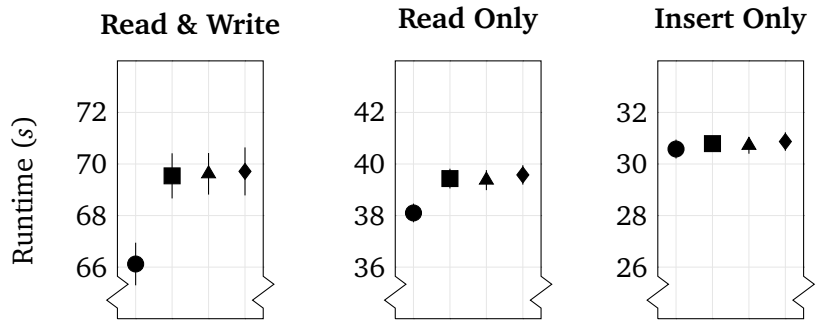

**Figure 5** Mean time required to run library reservation system workloads for the baseline (●), CapQL with no triggers (■), CapQL with triggers (▲), and ShillDB implementation (◆). 95 % confidence intervals are indicated by vertical bars (obscured by plotting symbols when intervals are small). Note: scale is consistent between plots but y-axes begin at different values.

**5.2 Performance Analysis**

The prototype implementations of ShillDB and CapQL focus on security, not performance. Nonetheless, we used the library reservation system as a performance benchmark to verify that the performance overhead of using ShillDB is reasonable.

We explore performance via four different implementations. As a baseline, we implemented the reservation system using Racket's standard database library. To examine the overhead of using CapQL instead of the standard database library, we implemented two different versions of the server in Racket using CapQL. One uses a modified CapQL that does not install database triggers to enforce view constraints for updates or inserts, and the other implementation uses CapQL as described in section 4.1. The final implementation is the Racket/ShillDB implementation complete with contracts as described in section 5.1.

For each implementation, we considered three different workloads. The first workload consists of 1,500 requests that require both reading from and writing to the database (e. g., looking up books, adding new reservations, deleting existing reservations). The second workload consists of 750 requests that require only database reads. The third workload consists of 2,000 requests that require only insert operations.

We ran each implementation for each workload 50 times on a four core, 2.7 GHz i7 machine with 8 GB of RAM running macOS 10.12.6. Figure 5 displays the result.

First, observe that the slowdown for the ShillDB implementations is small for all three workloads: the largest slowdown was 5.43 % in the read & write workload.

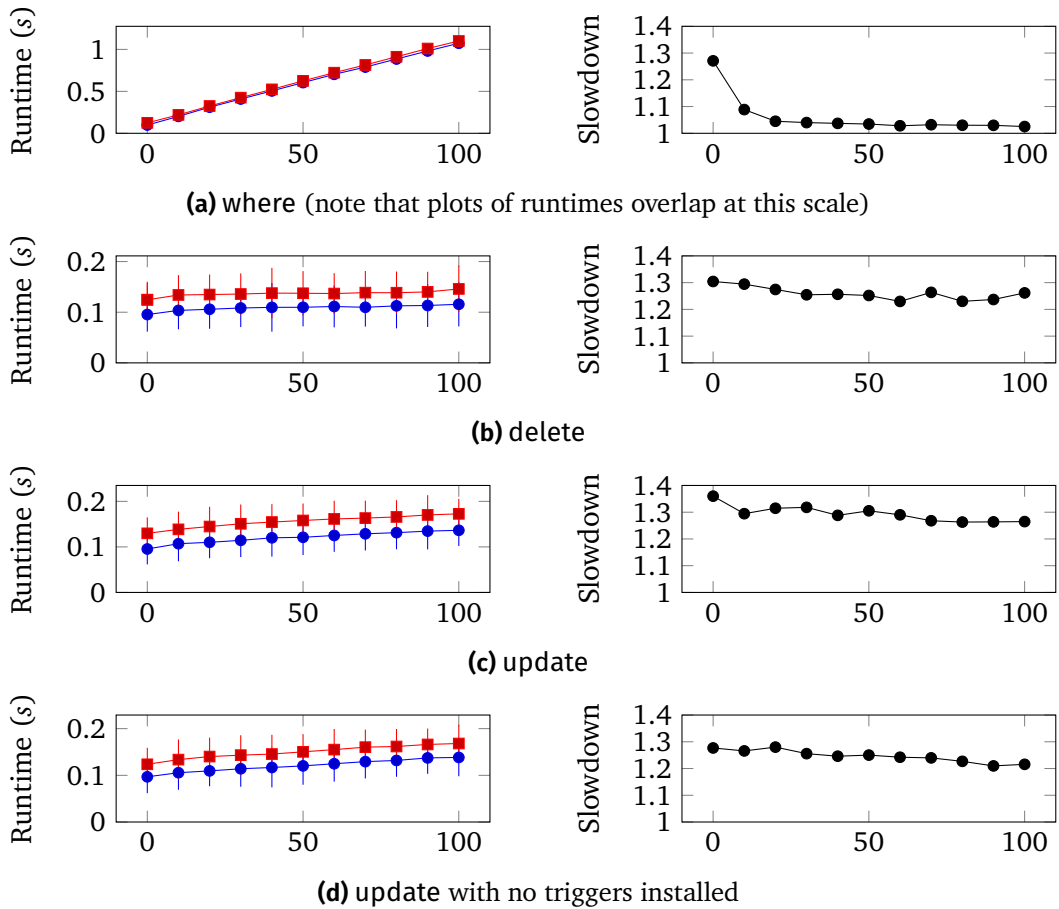

**Figure 6** Results for each CapQL benchmark. The left column shows mean runtime vs. query selectivity % for the baseline (●) and CapQL (■) implementations. Vertical bars indicate 95 % confidence intervals. The right column plots the slowdown of the CapQL implementation compared to the baseline vs. query selectivity %.

Second, note that the biggest difference in performance is between the baseline and CapQL implementation (we further explore the overhead of using CapQL below). Adding database triggers to enforce view constraints and adding ShillDB contracts both result in negligible slowdowns compared to the CapQL implementation. Finally, note that ShillDB and CapQL have comparatively large slowdowns in the read & write and read-only workloads (5.43 % and 3.87 % respectively) compared to the insert-only workload (0.95 %). The following section examines the performance characteristics of different CapQL operations to better explain this finding. Overall, these results suggest that future performance optimization ought to focus on CapQL.

### 5.2.1 CapQL Benchmarks

To better understand the overhead of database operations due to the implementation of CapQL, we evaluated small benchmarks using Racket's standard database library and CapQL operations. All benchmarks use a simple schema consisting of one table with two integer columns. The where benchmark performs a selection and a fetch

**Fine-Grained, Language-Based Access Control for Database-Backed Applications**

for a particular query selectivity (that is, the portion of rows that a query affects). The delete benchmark deletes a subset of rows. The update benchmark performs an arithmetic update on a subset of rows in the table. To measure the overhead of installing and executing the database triggers for views, the CapQL implementation of the update benchmark restricts the view before updating using a simple where clause. The insert benchmark inserts ten rows into the table. As in the update benchmark, the CapQL implementation restricts the view before inserting. For both the update and insert benchmarks, there is an accompanying version that uses a modified CapQL implementation that does not install database triggers.

We timed each benchmark 100 times against a table with 50,000 rows for a variety of selectivity values between 0 % and 100 % (except for the insert benchmarks for which there is no notion of selectivity). We ran the benchmarks using the same hardware as for the case study. Figure 6 shows the mean execution times with 95 % confidence intervals (left column) and the mean slowdown for the CapQL implementation compared to the baseline (right column). The mean slowdown for the insert benchmark (not shown in the figure) was 1.08× with triggers and 1.02× without triggers.

First, note that for read operations, the slowdown due to CapQL is negligible for large queries: the fixed overhead of using CapQL is dominated by the high cost of fetching results into memory. Second, while the slowdown trends downwards with increasing query size for update and delete operations, the overhead is still significant for large queries. This is not due to the overhead of checking triggers in updates because the trend is consistent for updates, for deletes (which do not install triggers), and for the modified update benchmark with no triggers. The cost of executing updates and deletes at the DBMS level in this case does not increase dramatically enough to dominate the overhead of CapQL, even for large queries. Finally, these results are consistent with the performance results from the library case study. Low selectivity deletes, updates, and fetches are the worst case for CapQL compared to the standard database library, whereas the overhead of insertions is small. This corroborates the result that the read & write workload (which consists primarily of low-selectivity deletes and fetches with some inserts) and the read workload (low-selectivity fetches only) performed much worse than the insert-only workload.

Finally, to understand why certain CapQL operations have such significant overhead compared to Racket's standard database interface, we used Racket's statistical profiler [21] to profile one of the CapQL microbenchmarks. We ran the where benchmark with 0 % query selectivity 10,000 times, and the profiler collected samples to estimate the execution costs for different functions called during the benchmark. The results suggest that a significant portion of running time (about 1/3) was spent parsing and validating the WHERE clause argument supplied to the where operation. This helps explain the results of the microbenchmarks, as where, update, and delete all must validate SQL expressions they are passed, while insert does not have any SQL expression arguments. Further, this suggests that future performance optimizations ought to focus on speeding up validation of SQL arguments or pushing more of this work to the DBMS (for example, by caching prepared queries or by using stored queries in the DBMS). Since the SQL-parsing functionality exists for usability, we could also provide an option to disable it.

3:20

## 6 Related Work

We overview three categories of related work in language-based security. For a comprehensive review of access control at the DBMS level, see Bertino, Ghinita, and Kamra [1].

**Language Support for Principle of Least Privilege** Other systems have also used language-level capabilities to support POLP. The E programming language [17] is an *object capability language*, where object references are treated as capabilities to invoke operations on that object. Passing an object reference provides unattenuated access to the underlying object, but it is possible to protect sensitive behaviors of objects by wrapping them in proxy objects. Both SHILLDB and Shill use contracts to express and enforce these *access abstractions* at component interfaces. CapDesk [26] provides support to launch applications written in E with limited authority. Melicher, Shi, Potanin, and Aldrich [12] propose a module system where modules are first-class capabilities: a module can be accessed only via a capability for that module, and capabilities for sensitive *resource modules* are obtained only as arguments to an ML-style functor [11].

Many research efforts seek to limit mainstream languages to capability-safe subsets to better support reasoning about and limiting the authority of programs. These approaches typically limit the API of the original language and restrict access to ambient authority. Examples include Joe-E [13] (a subset of Java), Emily [25] (a subset of OCaml), and Caja [15] (a subset of JavaScript). SHILLDB and Shill are capability-safe subsets of Racket. Specifying and verifying security policies in object-capability languages is still an active area of research [7]. However, since capability-safe languages enable security reasoning based on the flow of capabilities, informal security reasoning can be easier than in languages with ambient authority.

**Software Contracts for Security** Prior work also used contracts to enforce access control policies. Heidegger, Bieniusa, and Thiemann [10] introduce access permission contracts to restrict the fields that a method can access. Contracts have also been used to restrict how capabilities flow between components in object-capability languages [5]. Other contract systems allow enforcing more general policies. Moore, Dimoulas, Findler, Flatt, and Chong [18] use *authority environments* to manage rights within an execution context and *authorization contracts* to limit authority environments.

**Language-based Database Security** Caires, Pérez, Seco, Vieira, and Ferrão [2] propose a refinement type system that statically ensures programs adhere to database access control policies. UrFlow [3] also uses static policies, written as SQL queries, to enforce access control policies.

The refinement type system, UrFlow, and SHILLDB are different points in the design space of language-based database security. In the refinement type approach, programmers specify policies for each column of a table (for example, "you can see my profile picture only if you are my friend") and add annotations to aid the type checker. In UrFlow, programmers write policies using SQL queries that define what data a user can see (so the same photo policy would be written as a query fetching the profile pictures a user can view) and do not require additional annotations. In SHILLDB, access is controlled by database capabilities and further refined by policies expressed as contracts on capabilities. To express the same profile picture policy, a developer would write a contract allowing the picture column to be fetched only after doing a join to the current user's friends. All three of these approaches are declarative (i. e., developers write policies but not how to enforce them) but present different interfaces to express policies.

When considering the trade-off between ShillDB's run time validation of contracts and compile time approaches, we consider three main axes of comparison. For a general comparison between contract systems and other enforceable specification techniques, see Dimoulas, New, Findler, and Felleisen [6].

1. **Runtime performance overhead**: Performance overheads of run time checks for contracts are well-studied and are an active area of research [8]. While our benchmarks show that the contract checking overhead is reasonable in the artificial cases considered, the overhead may not be acceptable for all applications. Note that the checking cost from contracts is pay-as-you-go: removing contracts or making contracts less precise decreases the overhead.

2. **Annotation burden**: To use a compile-time technique, such as a type system, programmers must typically provide hints to the checker, such as type annotations, often to significant portions of code. Compare this to Racket and SHILLDB where contracts can be added gradually, starting with no contracts at all. Further, these contracts are expressed in terms of the normal program semantics and thus present a lower barrier to entry, while usually the annotations for compile-time techniques are expressed in terms of some formal logic.

3. **Expressiveness**: At run time, contracts can enforce properties that cannot be easily checked statically, such as properties that depend on the dynamic contents of a database table. However, run time validation of contracts cannot prove safety properties of a program in general (e. g., "this function never writes a null value to the database") — it can only prove that a specific execution of the program does not violate the contract. Finally, having run time contract failures halt program execution may be unacceptable in some settings.

Other approaches have considered the problem of *information flow* instead of access control. Information flow is able to address the propagation of sensitive information, in addition to access to data. SeLINQ [23] and UrFlow [3] use static types to reject queries that violate information flow policies. Jacqueline [30] performs dynamic query rewriting to enforce information flow policies (like SHILLDB does for some access control policies). Jacqueline uses a *policy-agnostic* approach in which information flow policies are separated from the rest of the application code. Similarly in SHILLDB, access control policies are pushed to function interfaces.

## 7 Conclusion

SHILLDB provides language-level support for running database-backed applications with least privilege and enforcing database access control policies. SHILLDB is capability safe and ensures that programs access database resources based only on the capabilities they are given. SHILLDB provides contract combinators tailored for writing fine-grained specifications for the use of database views, and the SHILLDB runtime enforces these specifications. This allows pushing access control policies to program interfaces, making it easy to inspect or modify policies without looking at the program's implementation.

**Acknowledgements**   We thank James Mickens and Margo Seltzer for their helpful comments.

**Fine-Grained, Language-Based Access Control for Database-Backed Applications**

Barbara, CA, USA: ACM, 2016, pages 631–647. ISBN: 978-1-4503-4261-2. DOI: 10.1145/2908080.2908098.

## A   SHILLDB Contract Reference

### A.1   SHILLDB Contract Syntax

**view/c:**   The general form of the view/c contract is:

```
1  (view/c privilege ...)
2
3       privilege = simple-privilege
4                 | [simple-privilege modifier ...]
5
6       simple-privilege = +fetch
7                        | +update
8                        | +delete
9                        | +insert
10                       | +where
11                       | +select
12                       | +aggregate
```

Which produces a contract for a view capability with only the privileges listed, where privilege ... indicates one or more privilege expressions and modifier ... indicates one or more modifier expressions (appendix A.2 has a listing of available modifiers). As a special case, view/c on its own (with no privileges) checks only that a value is a SHILLDB view capability and does not wrap the capability in a proxy, effectively conferring all privileges if the capability was not previously wrapped in a more restrictive contract.

**->/join:**   The general form of the ->/join contract combinator is:

```
1  (->/join (group-definition ...) dom-expr ... range)
2
3       group-definition = [identifier join-modifier ...]
4
5       dom-expr = ctc
6                | [ctc #:groups group-identifier ...]
```

Which produces a contract on a function which, when applied, places arguments in the join groups defined in the contract. Above, ctc is any valid contract (including standard Racket contracts), range is any valid contract or any, join-modifier ... is one or more modifier expressions taken from the list of valid join modifiers in appendix A.2, identifier is a valid Racket identifier, and group-identifier ... is one or more of the identifiers defined in the group-definition portion of the same contract. The contracts in dom-expr are contracts on the function arguments and range is the contract on the result (analogous to the structure of the -> contract combinator in Racket).

## A.2 SHILLDB Privilege Modifiers

| Privileges | Modifier | Description | Example |
|---|---|---|---|
| `+fetch` `+update` `+delete` `+insert` | `#:restrict` | Provides a restricted window into the view for an operation based on the given view to view function. | `#:restrict (λ (v) (where v "id < 10"))` |
| `+aggregate` | `#:having` | Filters out any groups in the resulting view that do not satisfy the given HAVING clause. | `#:having "COUNT(*) > 10"` |
| | `#:aggrs` | Rejects any aggregation query that contains an aggregation function other than those listed. | `#:aggrs "MIN, MAX"` |
| | `#:with` | Specifies what contract the view should derive after an aggregation. | `#:with (view/c +fetch)` |
| `+join` | `#:pre` | Rejects any joins that do not satisfy the given predicate over tables and join condition. | `#:pre valid-foreign-key?` |
| | `#:post` | Applies the given view to view function to the result of joins. | `#:post (λ (v) (select v "id"))` |
| | `#:with` | Specifies what contract the view should derive after a join. | `#:with (view/c +fetch)` |
| `+where` | `#:prohibit` | Specifies columns that cannot be referred to in WHERE clauses. | `#:prohibit "gpa"` |
| `+select` | --- | --- | --- |

**Figure 7** Privileges and modifiers in SHILLDB. Modifiers can be used to refine what operations a particular privilege permits. To come up with the available modifiers, we identified cases where it was desirable to have more nuanced restrictions on operations, but we do not claim that this set of modifiers is complete in their expressiveness. In the example, valid-foreign-key? in the example used for #:pre is a function that takes two views and a WHERE clause and returns true just when either of the tables corresponding to the views has a foreign key field for the other table (based on schema information stored in the view capability) and the WHERE clause represents an equijoin on that key.

## B  Library Reservation Backend

### B.1  Schema

| cardholders | | |
|---|---|---|
| *card_id* | *firstname* | *lastname* |
| 1 | Steve | Martin |
| 2 | Richard | Pryor |

| authors | | |
|---|---|---|
| *author_id* | *firstname* | *lastname* |
| 1 | Trevor | Noah |
| 2 | Tina | Fey |

| books | | | |
|---|---|---|---|
| *book_id* | *author* | *title* | *copies* |
| 1 | 1 | Born a Crime | 4 |
| 2 | 2 | Bossypants | 6 |

| reservations | | |
|---|---|---|
| *r_id* | *book* | *cardholder_id* |
| 1 | 2 | 2 |
| 2 | 1 | 2 |

■ **Figure 8** Schema and example data used by the library reservation backend.

### B.2  Contracts on Endpoints

Figure 9 shows the definitions of the library reservation server endpoints along with contracts. The definitions use the **define/contract** macro which is similar to **define** except that it also takes a contract and attaches that contract to the bound value.

```
1  (define/contract (reserve book v-reservation)
2    (-> string?
3        (view/c [+insert #:restrict (lambda (v) (where v (sqlformat "cardholder_id = $1"
            ↪ (current-user))))])
4        any/c)
5    #| Implementation... |#)
6
7  (define/contract (my-reservations v-reservations v-books v-authors)
8    (-> (view/c
9          [+fetch #:restrict (lambda (v) (where v (sqlformat "cardholder_id = $1" (current-user))))]
10         +join +where +select)
11        (view/c +join +fetch +select +where)
12        (view/c +join +fetch +select +where)
13        string?)
14    #| Implementation... |#)
15
16 (define/contract (remove-reservation rid v-reservations)
17   (-> string?
18       (view/c +where
19         [+delete #:restrict (lambda (v) (where v (sqlformat "cardholder_id = $1" (current-user))))])
20       any/c)
21   #| Implementation... |#)
22
23 (define/contract (search-author fname lname v-authors v-books)
24   (-> string?
25       string?
26       (view/c +fetch +join +select +where)
27       (view/c +fetch +join +select +where)
28       string?)
29   #| Implementation... |#)
30
31 (define/contract (num-reservations book-id v-reservations)
32   (-> string?
33       (view/c [+aggregate #:with (view/c +fetch)] +where)
34       string?)
35   #| Implementation... |#)
```

■ **Figure 9** Library reservation backend endpoints with contracts. The implementations are omitted for brevity.

Fine-Grained, Language-Based Access Control for Database-Backed Applications

**About the authors**

**Ezra Zigmond** is a software engineer in Palo Alto, California. He received a bachelor's degree in Computer Science from Harvard College where his research interests included language-based security and program synthesis. Contact him at ezigmond@acm.org.

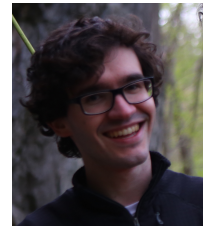

**Stephen Chong** is a Gordon McKay Professor of Computer Science in the Harvard John A. Paulson School of Engineering and Applied Sciences. Steve's research focuses on programming languages, information security, and the intersection of these two areas. He is the recipient of an NSF CAREER award, an AFOSR Young Investigator award, and a Sloan Research Fellowship. He received a PhD from Cornell University, and a bachelor's degree from Victoria University of Wellington, New Zealand. Contact him at chong@seas.harvard.edu.

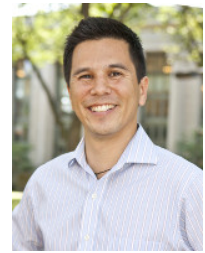

**Christos Dimoulas** is an Assistant Professor of Computer Science at Northwestern University. Christos' research focuses on the design and semantics of programming languages. Specifically, his goal is to develop programming languages technology that facilitates the construction of secure and robust component-based software systems. He received a PhD from the College of Computer and Information Science at Northeastern University. Contact him at chrdimo@northwestern.edu.

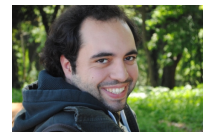

**Scott Moore** is a Principal Scientist at Galois. Scott's research focuses on applying programming language theory and analysis techniques to understand, detect, and repair exploitable vulnerabilities in programs. Previously, he was a postdoctoral fellow at Harvard University where he worked on using software contracts to express and enforce access control policies. He received a PhD from Harvard University, and a B.S. and M.S. in computer science at the University of Texas at Dallas. Contact him at scott@galois.com.

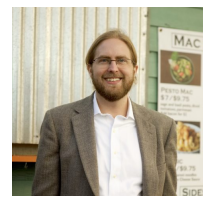

3:30